\documentclass[journal]{IEEEtran}
\usepackage[cmex10]{amsmath}
\usepackage[noadjust]{cite}

\usepackage{url}
\usepackage{color}

\usepackage{amssymb}
\usepackage{path}
\usepackage{verbatim}
\usepackage{algorithm}
\usepackage{framed}
\usepackage{algpseudocode}

\usepackage{color}
\usepackage{amsmath}
\usepackage{cases}
\usepackage{theorem}
\usepackage[export]{adjustbox}

\usepackage{graphicx}

\usepackage{pgfplots}
\usepackage{pgfplotstable}
\pgfplotsset{compat=newest}

{ \theorembodyfont{\normalfont} 

}

\newcounter{enumctr}

\DeclareFontFamily{U}{mathx}{\hyphenchar\font45}
\DeclareFontShape{U}{mathx}{m}{n}{<-> mathx10}{}
\DeclareSymbolFont{mathx}{U}{mathx}{m}{n}
\DeclareMathAccent{\widebar}{0}{mathx}{"73}

\begin{document}
%
\title{A Fair and Privacy-Aware EV Discharging Strategy using Decentralized Whale Optimization Algorithm for Minimizing Cost of EVs and the EV Aggregator}

\author{Yingqi Gu 
	and Mingming Liu	
	\thanks{Y. Gu is with the Insight Centre for Data Analytics, Dublin City University, Dublin, Ireland (e-mail: yingqi.gu@dcu.ie)}
	
	\thanks{M. Liu (\textit{corresponding author \& joint first author}) is with the School of Electronic Engineering, Dublin City University, Dublin, Ireland (e-mail: mingming.liu@dcu.ie)}

   \thanks{The authors acknowledge the funding support from the School of Electronic Engineering and the Entwine Research Centre at Dublin City University to conduct this research work. }}

\markboth{Journal of \LaTeX\ Class Files,~Vol.~6, No.~1, January~2007}%
{Shell \MakeLowercase{\textit{et al.}}: Bare Demo of IEEEtran.cls for Journals}

\maketitle

\begin{abstract}
	

A key motivation to fasten roll-out of electric vehicles (EVs) to the market is to implement Vehicle-to-Grid (V2G) functionalities. With V2G in place, EV owners can have an extra freedom to interact their battery energy with power grids, namely by selling their energy to the grid when their EVs are not in use. On the other hand, EV aggregators and utility companies can leverage the flexibility of the collected energy to implement various ancillary services to the grids, which may significantly reduce costs of, for instance, running spinning reserve of traditional power plants on the grid side. However, this extra freedom also poses practical challenges in terms of how to devise a discharging strategy for a group of EVs that is fair and in some sense optimal. In this paper, we present a new design of EV discharging strategy in a typical V2G energy trading framework whilst leveraging the whale optimization algorithm in a decentralized manner, a metaheuristic algorithm that has been shown effective in solving large-scale centralized optimization problems. We demonstrate that by using simple ideas of data shuffling and aggregation, one can design an EV discharging strategy in a fair, optimal and privacy-aware manner, where the privacy refers to the fact that no critical information of EVs should be exchanged with the EV aggregator, and vice versa. The fairness implies that a common discharge rate needs to be sought for all EVs so that no one gets better benefits than others in the same V2G programme. Simulation results are presented to illustrate the efficacy of our proposed system. 
	
\end{abstract}

\begin{IEEEkeywords}
	Electric Vehicles, Vehicle-to-Grid,  Decentralized Optimization, Whale Optimization Algorithm
\end{IEEEkeywords}

\IEEEpeerreviewmaketitle

\section{Introduction}

In recent years, there has been an increasing interest in providing Vehicle-to-Grid (V2G) as a service to users of Electric Vehicles (EVs) \cite{liu2013opportunities, sortomme2011optimal, tan2016integration}. The key concept of V2G relies on the fact that it allows bidirectional power flow between EVs and power grids, usually with an EV aggregator placed in the middle acting as an agent for energy trading in the electricity market \cite{bessa2010role, escudero2012fair}. The overarching goal for the operation of such a V2G system is not only to maximize the benefits for the EV aggregator but also to optimize the benefits for the EV owners so that enough EVs can be encouraged to participate into a V2G programme \cite{richardson2013encouraging}. For instance, an EV user may feel very reluctant to use V2G as a service if more energy has to be dispatched from the vehicle than its expected revenue that can be received in the end. Thus, it becomes a practical challenge to find out a balanced V2G strategy not only to maximize the benefits of an EV aggregator but also to maximize the benefits of all participated EVs in a relatively fair manner. 

Hitherto, a large body of works can be found in literature for providing ancillary services to the grids through V2G, and most of which have a strong focus on the frequency regulation service, see \cite{han2010development, kempton2008test, liu2013decentralized, zhong2014coordinated, escudero2012fair, shi2011real, sun2014real, ko2017effect} for some selected works in this direction. More specifically, from a fair design perspective, the paper \cite{escudero2012fair} discussed a set of approaches, including water-filling, state-dependent utility and SOC variance minimization, to regulate V2G energy delivery of EVs for the grid frequency regulation service according to different specific fairness criteria. In \cite{peng2017optimal}, an optimal dispatching strategy was presented for a V2G aggregator participating in supplementary frequency regulation while considering EV driving demand and the benefits of an aggregator, where a fair regulation power allocation module was built to avoid over-discharging of EVs. Furthermore, a real-time welfare-maximizing regulation allocation algorithm was proposed in \cite{sun2014real} in order to fairly allocate the regulation power capacity among the EVs for the aggregator. In \cite{xie2016fair}, an adaptive dynamic programming method was proposed to maximize the long-term fairness of EVs. The proposed method has been implemented in a way that EVs with high State-of-Charge (SOC) are chosen to discharge energy for load shaving task while the EVs with high contributions can have high priority to be charged afterwards. In \cite{han2010development}, an optimal control strategy using dynamic programming was adopted for the V2G frequency regulation services. In particular, the authors assumed a fair distribution of the regulation request to the pertaining vehicles in the study. 

Apart from using V2G for frequency regulation services, other approaches have also been found in order to manage the power flow in a fair and decentralized manner. For instance, Additive Increase Multiplicative Decrease (AIMD) and other network inspired methodologies were adopted in \cite{liu2015enhanced} and \cite{liu2013fair}, in order to seek fair allocation of EV power flows while considering a set of specific power system constraints. In particular, \cite{liu2013fair} also considered a proportional fairness based algorithm inspired by a distributed price feedback mechanism.  Furthermore, a fair V2G discharging strategy was proposed in \cite{liu2014optimal} where a utility optimization problem has been solved by taking account of the benefits of utility companies and EV users' inconvenience of using V2G in a microgrid scenario. More specifically, the fairness criteria refers to the fact that same amount of power needs to be taken from all EVs in the V2G programme to avoid some EV owners' having more benefits than other users. Finally, we note that some works also have a strong focus on the privacy-preserving perspective in V2G, where the main idea was to not reveal any sensitive information during the information processing, coordination and communication exchange between EVs and a central computing node, e.g. an EV aggregator, using V2G; typically these information may include an EV user's personal ID, an EV's location information, as well as payment and billing information \cite{han2016privacy}. To address these issues, decentralized based approaches have been more preferable in V2G practices, see \cite{gao2018blockchain, garcia2016multi, xing2015decentralized} for some recent use cases. Finally, a comprehensive comparison between different centralized and decentralized based optimization techniques for EV charging/discharging control has been reported in papers \cite{vaya2012centralized, fang2011smart, staudt2018decentralized, hoang2017charging}.

Along this line, our objective in this paper is to design a V2G programme for the mutual benefits of EV users and the EV aggregator by jointly solving a constrained consensus optimization problem, where each party can only get access to its own part of the objective function (privacy-preserving). In particular, we shall assume that each part of the objective function is treated (encapsulated) as the ``black-box'' model and only limited information can be obtained from the model, e.g. no derivative information can be retrieved. In fact,  such an assumption is not uncommon in current practices as more data have now been processed and trained in a non-local environment, e.g. cloud, and thus even though the model is visible to the model creator it may not be fully visible or explainable to end-users. In this regard, a centralized-based heuristic algorithm, e.g.  Centralized WOA (CWOA) \footnote{In the following, we shall refer the WOA proposed in \cite{mirjalili2016whale} as CWOA.}, may be plausible for an optimal solution, but it does not take into account users' privacy concerns as sensitive information may be collected from different users to carry out this computation process. Also, a centralized based solution may not easily handle hard constraints in an optimization problem, especially in terms of the consensus constraints of our interest here. 

Thus, our contribution of this paper is to propose a fair (consensus) and privacy-preserving power management mechanism by including the following features in a V2G programme. 

\begin{itemize}
	\item[A.]   A privacy-aware communication mechanism which enables various information to be safely exchanged among EV users and the EV aggregator in a V2G programme.  
    \item[B.]   A system model which captures the modelling procedures of costs for both EVs and the EV aggregator in V2G.
	\item[C.]  An effective and efficient optimization algorithm that can deal with the ``black-box'' models for optimization.
	\item[D.]  A practical system architecture that can integrate the three parts, i.e., A, B and C, together.
\end{itemize}

The remainder of the paper is organized as follows. Section \ref{model} presents the system model for the V2G power dispatch problem. Section \ref{algorithm} reviews the existing algorithms and proposes the system implementation steps using the decentralized WOA. Section \ref{Simulation} demonstrates our simulation setup and presents our simulation results. Section \ref{conclusion} concludes the paper. Finally, Section \ref{future} gives a remark on the limitations of our current approach and outlines some thoughts for future work.

\section{System Model}\label{model}

\subsection{System Set-up} \label{model_des}

We consider a scenario where a number of EVs are plugged-in a large parking area managed by an EV aggregator. In particular, some EVs can opt-in a V2G programme, and such EVs can discharge certain amount of energy to the grid for some economic revenues. In reality, these EV owners can be local residents who work nearby in the parking area. The EVs may have already been fully charged at home, e.g. by using home solar PhotoVoltatic (PV) panels or distributed small wind turbines, before travelling to the parking area. An EV as such can reserve only a small amount of energy for travelling back home, and trade in most of the energy stored in its battery pack for monetary benefits. With this in place, an EV aggregator can leverage the collected battery energy from EVs to provide ancillary services to the main power grids. We shall require that the designed discharge rate is consistent for all EVs to avoid having some EVs getting more benefits than others. Finally, we note that although an EV aggregator may also provide an EV charging service for many parked EVs, it is not our main focus in this work as our targeted EV users are those mostly interested in making revenues from the V2G service. Thus, we shall ignore the EV charging part in our system model design.

We now formulate the EV discharging problem as follows.  Let $N$ be the maximum number of EVs participated in the V2G programme during a certain period of time, e.g. during peak time when grid needs most energy regulation. Define the set $\underline{\textrm{N}}: = \left\lbrace 1,2,3, \dots, N \right\rbrace$ for indexing total EVs in the programme, and also the set $\underline{\textrm{N}}(t)$, which is a subset of $\underline{\textrm{N}}$, for indexing all available EVs in the programme at time $t$, that is some EVs may become unavailable due to finishing the V2G programme or leaving the parking lot earlier.  Let $c_i(t)$ be the discharge rate of the $i$'th vehicle at time $t$, with $c_{\textrm{min}}^i$ and $c_{\textrm{max}}^i$ defined as the minimum and maximum discharge rate of the vehicle, respectively.  In addition, we denote $SOC_i(t)$ the state-of-charge of the $i$'th vehicle at time $t$, and let $SOC_{\textrm{min}}^i$ be the minimum state-of-charge of the $i$'th vehicle that an EV user can just accept in the V2G programme. In other words, if $SOC_i(t) < SOC_{\textrm{min}}^i$ then significant inconvenience will be imposed to an EV user, and thus $c_i(t)$ will be automatically set to 0 in this situation. In practice, for example, an EV user may set $SOC_{\textrm{min}}^i$ to 10\% if the user's home is very close to the parking lot, $20\%$ if the user's home is relatively far from the parking lot.

During parking, each EV is connected to a V2G discharge point. We assume that each discharge point can communicate to a central computing server/node bidirectionally, and each discharge point can also send a broadcast information to other discharge points in the same parking area. We note that such a communication requirement can be easily satisfied through, e.g., powerline communication infrastructures, in a practical V2G scenario. However, any capable IoT device that has Wifi/3G/LTE/5G capability on the discharge point can also be used as an alternative. In any case, the communication links are required so that information can be exchanged among EV discharge points and a central computing server to jointly determine the optimal V2G power dispatch rate.

Finally, we assume that each EV $i$ is associated with a cost function, $f_i(c_i(t))$, which quantitatively characterises the net cost when the $i$'th EV is selling $c_i(t)$ power to the EV aggregator through V2G. Here, the net cost refers to the fact that although an EV can receive monetary benefits by selling its energy, there will still be an operational cost incurred as a certain level of inconvenience will be imposed to such an EV, including discharge rate related battery wear and degradation cost. Similarly, when an EV aggregator is buying energy from EVs, the EV aggregator will benefit from the collected energy from all EVs using V2G. However there will also be an operational cost incurred which includes the energy cost paid to the EV users as well as the infrastructure maintenance cost during this process. We denote $Agg(t)$ the net cost function of the EV aggregator in delivering the V2G service for EV users at time $t$, which is essentially a function in terms of $\sum_i c_i(t)$.
	
Our objective in this paper is to solve the following optimization problem:

\begin{equation} \label{eq:opt}
\begin{gathered}
\underset{c_i(t)}{\min} \quad
\sum_{i \in \underline{\textrm{N}}(t)} f_i(c_i(t)) + Agg(t)  \\
{\text{s.t.}} ~
c_i(t) = c_j(t), ~ \forall i \neq j \in \underline{\textrm{N}}(t) \\
c_{\textrm{min}}^i \leq c_i(t) \leq c_{\textrm{max}}^i, ~\forall i \in \underline{\textrm{N}}(t) \\
SOC_{\textrm{min}}^i \leq SOC_i(t) \leq 100\%, ~\forall i \in \underline{\textrm{N}}(t)
\end{gathered}
\end{equation}

\noindent where we wish to find an optimal consensus solution $c^*(t)$ for all EVs so that the net cost for all EVs and the EV aggregator can both be minimized. In order to solve this problem, we first explicitly model the cost function $f_i(c_i(t))$ for EV $i$ and then we model the cost function $Agg(t)$ for an EV aggregator.




\subsection{Cost Function for EVs}

Now we model the net cost function $f_i$ for the $i$'th EV. Specifically, we shall take account of three factors: 1. monetary benefits when the EV sells its energy at discharge rate $c_i(t)$; 2. discharge rate related battery degradation costs; and 3. other operational costs, e.g. cable wear out and various relevant service charges. For simplicity, we assume that the discharge rate related degradation cost is dominant compared to other V2G related operational costs. To begin with, we denote $u_i(c_i(t))$ the monetary benefits for the EV selling its power at discharge rate $c_i(t)$. Let $p(t)$ be the unit price of V2G power defined by the aggregator at time $t$, then we have: 

\begin{equation}\label{ev_moneypart}
u_i(c_i(t)) = p(t) c_i(t).
\end{equation} 

Note that the equation \eqref{ev_moneypart} simply indicates that the more energy is provided by the EV, the better monetary benefit can be obtained by the EV. Next, let $d_i(c_i(t))$ be the degradation cost when $c_i(t)$ power is drawn from the EV $i$. In literature, this degradation cost can be modelled using an quadratic function in the following form \cite{ma2015distributed}:

\begin{equation} \label{ev_degradationpart}
d_i(c_i(t)) = \alpha_i c_i(t)^2 + \beta_i c_i(t) + \gamma_i, 
\end{equation}
where $\alpha_i$, $\beta_i$ and $\gamma_i$ are all parameters. Finally, let $o_i(.)$ be the function which models other related operational costs of the EV in V2G. Since the function is weakly correlated to the discharge rate, we assume that the cost is a constant value across the range of all feasible discharge rates of the $i$'th EV. For simplicity, we shall use $o_i$ to model this lumped cost in the following context. With all these factors, we now define

\begin{equation} \label{ev_netcost}
f_i(c_i(t), p(t)) = d_i(c_i(t)) + o_i - u_i(c_i(t))
\end{equation}
as the net cost for the EV $i$ delivering its power at the rate $c_i(t)$ using V2G. We note that in most real-world cases, the electricity price per unit $p(t)$ can be consistent during a period of time in a day, e.g. peak-time tariff or off-peak time tariff. Given this viewpoint, our original cost function $f_i(c_i(t), p(t))$ can be equivalently described as $f_i(c_i(t); p)$, i.e. the function which is now parametrised by the fixed unit price $p$. In other words, for a given price $p$, $f_i$ is a function of single variable $c_i(t)$. This new way of describing $f_i(c_i(t), p(t))$ is important from an EV user's point of view as now the presented cost function is able to capture a user's discharging behaviour in V2G. It is also important to mention that the function contains some sensitive and private information including regular maintenance costs and other performance characteristic information of the EV. For this purpose, the function $f_i(c_i(t); p)$ should be \textit{strictly private} to the EV user in the V2G programme, i.e. an EV aggregator should not be given a permission to access this information in any associated computing process.

\subsection{Cost Function for an EV Aggregator}

\begin{figure*}[htbp]
	\begin{center}
		{\includegraphics[width=0.8\textwidth]{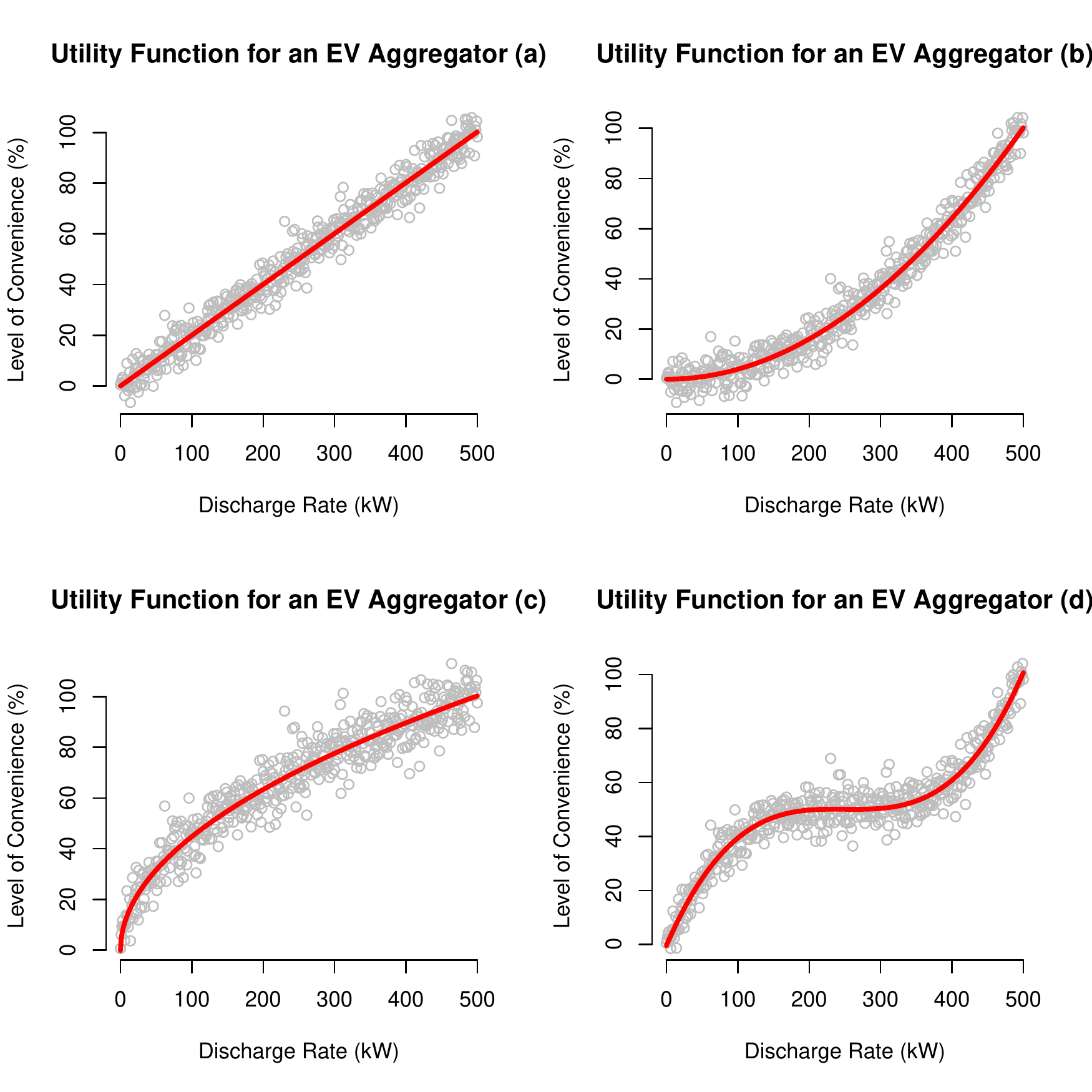}}
		\caption{Some illustrative examples for fitting an EV aggregator' utility function (100\% as the highest level) with respect to the overall V2G discharged power from EVs. The grey circles indicate an EV aggregator's input data points, and the red curves are fitted curves using different machine learning models. Different possibilities exist for fitting the curves as shown in subplots (a), (b), (c) and (d) given different set of data. In our context, these curves are private to the EV aggregator. The figure was generated in R. }
		\label{example_ml}
	\end{center}
\end{figure*}

In this section, we model the cost function $Agg(t)$ for an EV aggregator. Similar to what we have modelled for the EV cost functions, we consider two main factors for modelling the net cost incurred by an EV aggregator, including 1. the potential benefits that the collected V2G power can be leveraged by the EV aggregator in the energy market; and 2. the monetary cost of an EV aggregator for sourcing the V2G power from EV users. Specifically, let $U(t)$ be the utility that an aggregator can achieve in V2G at time $t$. Clearly, $U(t)$ is a function of $c_i(t)$ because at a given time slot $t$ the utility value $U(t)$ depends on the V2G power collected from all EVs $\sum_i c_i(t)$. We note that $U(t)$ essentially captures the convenience that an EV aggregator can benefit from the collected V2G power for various activities in the power system operations, such as frequency regulation and peak load compensation. Due to this diversity, how best to model $U(t)$ is still an open issue \cite{fan2012distributed}. For instance, a convex/concave utility function may be the best fit for a regulation service, but a non-convex curve could also be used to best model the efficacy of another grid regulation service. For simplicity of our model interpretation and numerical evaluation, we shall assume that a logarithmic function can be applied to model the beneficial part for an EV aggregator in this work, i.e. 

\begin{equation} \label{aggutility}
U(t) =  \omega \log (\sum_i c_i(t) + 1). 
\end{equation}

\noindent where $\omega$ is a parameter which can be used to reflect the level of convenience for an EV aggregator. The ``$+1$'' in \eqref{aggutility} is to allow the calculation when no V2G power is delivered from EVs at time $t$, i.e. $\sum_i c_i(t)=0$, which also indicates that no convenience will lead to the EV aggregator when no V2G power is dispatched from EVs. In reality, modelling of $U(t)$ can be done through machine learning and many neural network based approaches provided that enough historical datasets can be collected from an EV aggregator. Some illustrative examples for the potential choices of $U$ are presented in Fig. \ref{example_ml}. As shown in this figure, an EV aggregator needs to provide their data points indicated by grey circles for model training purposes. We shall ignore further discussions on this point as there are plenty of methods can be used for this application and it is beyond the scope of this paper. 

Concerning the monetary cost for an EV aggregator, instead of using a unit price signal $p(t)$, here we interpret this cost as the energy generation cost from the perspective of energy exchange in the electricity market. In particular, we see the procurement of V2G power as an energy generation process where the energy is not directly generated by physical power plants, e.g. wind power plants, Combined Heat and Power (CHP), PhotoVoltatic (PV) plants, but in a similar alternative manner. Namely, the incurred cost can be regarded as if the same amount power can be sourced from other physical power plants from an operational perspective. In this regard, it is worth noting that the quadratic functions have been commonly adopted for modelling the generation costs of power plants in the literature, see \cite{crisostomi2014plug, caldon2004optimal, parisio2011mixed}. With this in mind, we mathematically present the cost for the EV aggregator $M(t)$ as follows:



\begin{equation} \label{aggcost}
M(t) = a (\sum_i \eta_i c_i(t))^2 + b (\sum_i \eta_i c_i(t)) + c  
\end{equation}  

\noindent where $a, b, c$ are coefficients of the generation cost for a power plant, and $\eta_i$ denotes the energy inversion efficiency from DC to AC at each discharging point for the $i$'th EV, which implies that only $\eta_i c_i(t)$ amount of (AC) power will be eventually delivered to the grid. In practices, the efficiency factor $\eta_i$ can be modelled as a constant value depending on the specific types of vehicle models \cite{garcia2016multi} connected to the V2G discharge point. For instance, the energy conversion efficiency for a Nissan Leaf Plug-in EV can be 86.4\% as reported in \cite{garcia2016multi}. Since $a$ is usually a positive constant value, and thus the resulting function $M(t)$ is a quadratic function with convexity.  Combining both \eqref{aggutility} and \eqref{aggcost},  the net cost function $Agg(t)$ of an EV aggregator can be formulated as:  

\begin{equation} \label{aggnetcost}
Agg(t) = M(t) - U(t)  
\end{equation}  

\noindent \textbf{Comment: } We note that the parameters for cost functions in \eqref{ev_netcost} and \eqref{aggnetcost} can be well estimated using a data-driven approach, where data comes from a user's historical observations when using the V2G
programme. In particular, we assume that these cost functions are usually encapsulated using learning models and users can make real time predictions of their costs by invoking  such as REST API calls (as models can be deployed off-site). In this regard, other details of the model including mathematical expression of the fitted function and derivative related information of the cost function are usually implicit to the users. Thus, we assume that both cost functions $f_i$ and $Agg$ are \textit{``black-box''} models and only the function value can be evaluated but with a cost, e.g. communication \& API fees.

\section{Algorithms and Implementations} \label{algorithm}

In fact, the optimization problem \eqref{eq:opt} can be easily solved in a centralized optimization framework using programming techniques. However, a centralized based algorithm is usually implemented in a batch manner, typically by an optimizer located at the central computing node facilitated at the EV aggregator.  In this way, it is usually required that both cost functions $f_i$ and $Agg$ to be exposed to a central node, which may be impossible for EV users as sensitive information can be contained or factored into the cost function $f_i$. When a centralized algorithm is implemented and converged at the central node, the optimal solution will be communicated to each EV for further handling. In contrast, a decentralized based algorithm is designed to be implemented at a computing unit closer to the EVs, e.g. an edge node. In this set-up, each EV may still communicate limited information to its neighbouring networks and a computing node for feedback, but the overall goal is to find an optimal solution for EVs in a collaborative and privacy-preserving way. 
	
In this work, we are interested in the design of a decentralized algorithm that can be useful for addressing the problem \eqref{eq:opt}.  More specifically, a decentralized approach has several advantages over the centralized one, especially considering the perspectives of privacy preserving and agent actuations. In our problem, both EVs and the EV aggregator define their own cost functions which reflect their preference in choosing the optimal discharge rate in V2G for their benefits. Clearly, neither side is willing to share their own information to the other side. This problem becomes even more intractable given both cost functions are treated as ``black-box'' models.


\subsection{Existing Optimization Algorithms}

In the literature, many decentralized approaches have been proposed which are able to solve an optimization problem similar to \eqref{eq:opt}, see \cite{fan2012distributed, liu2013fair, liu2014optimal, boyd2004convex, liu2017stability, shi2014linear, huang2016consensus}. Among many others, Alternating Direction Method of Multipliers (ADMM) algorithm has been recently proposed as an evolution of other well-known optimization algorithms such as the method of multipliers and dual ascent \cite{ghadimi2014optimal, naoum2016smart}. As an alternative to ADMM-like algorithms, our key idea is to adopt a recently proposed Whale Optimization Algorithm (WOA) and to investigate how it can be adapted to solve the problem \eqref{eq:opt} in a decentralized framework while satisfying various requirements. Our key motivations for using WOA can be summarised as follows:  

\begin{itemize}
	\item \textit{Agent Actuation:} ADMM-like algorithms essentially decompose the central optimization problem to some sub-optimization problems which can be easily solved at the edge side. This requires each local agent being able to solve a ``smaller'' optimization problem through a local optimizer \cite{naoum2016smart}. In contrast to ADMM, the WOA based algorithm does not need such an ability, instead it requires each agent to simply follow pre-defined heuristic. \\
	\item \textit{Elasticity:} ADMM-like algorithms are mainly used to solve convex optimization problems by leveraging the derivative information of the objective function. Special designs are usually required to deal with complex non-convex objective functions. In contrast, the WOA-based algorithm is designed to deal with complicated objective functions without relying on derivative information of such functions. This feature makes the WOA-based algorithm an ideal tool to tackle with the optimization problem of our interest in this paper.
    \item \textit{Robustness:} WOA-based algorithm is essentially a heuristic algorithm and algorithm parameters are weakly correlated to the network dimension and the type of cost functions. As we shall see later, the WOA-based algorithm can converge effectively for our problem by changing values of different hyper-parameters. 
\end{itemize}

\subsection{Whale Optimization Algorithm} \label{WOA}

The WOA is inspired by the foraging behaviour in groups of humpback whales \footnote{Without explicit mentioning, we shall just call whales in our context.}. This special hunting method, also known as the bubble-net feeding method, is done by creating distinctive bubbles along a `9'-shaped spiral path \cite{mirjalili2016whale}, with an aim to encircle the prey and attack it (using bubble-net). At every time instance, there are only two actions could be done by a whale, namely either encircling a prey or using bubble-net to attack the prey, and both actions are done in a random manner. Moreover, there are two options for a whale when encircling a prey, that is a whale can either follow the current best whale, i.e. the one that is mostly close to the location of prey, or move towards a random whale's position. Readers of interest should refer to \cite{mirjalili2016whale} for more details of the centralized WOA and we shall ignore further discussion on the CWOA. Instead, we give some useful observations below. 

\begin{itemize}	
	\item CWOA is able to solve both constrained and unconstrained optimization problems. For constrained optimization problems, a fitness function usually includes a penalty term to reduce the search space. The simplest choice of the penalty function is the death penalty \cite{mirjalili2016whale}, which simply assigns a large value (for minimization problems) to the objective function in the case that constraints cannot be satisfied. In our problem, this implies that a search agent cannot be the optimal agent (solution) if the position of such an agent is not in the consensus form, which is a hard constraint for CWOA.  
	\item CWOA intends to find out the best search agent based on a fitness function which depends on the states of all search agents. In our context, this fitness function requires the information of all cost functions of EVs and the EV aggregator, which addresses the concern on ``black-box'' models but not on the privacy-preserving aspect. 
\end{itemize}

\subsection{Decentralized WOA and Proposed System Implementations}

\begin{figure*}[htbp]
	\begin{center}
		{\includegraphics[width=0.8\textwidth, height=4in]{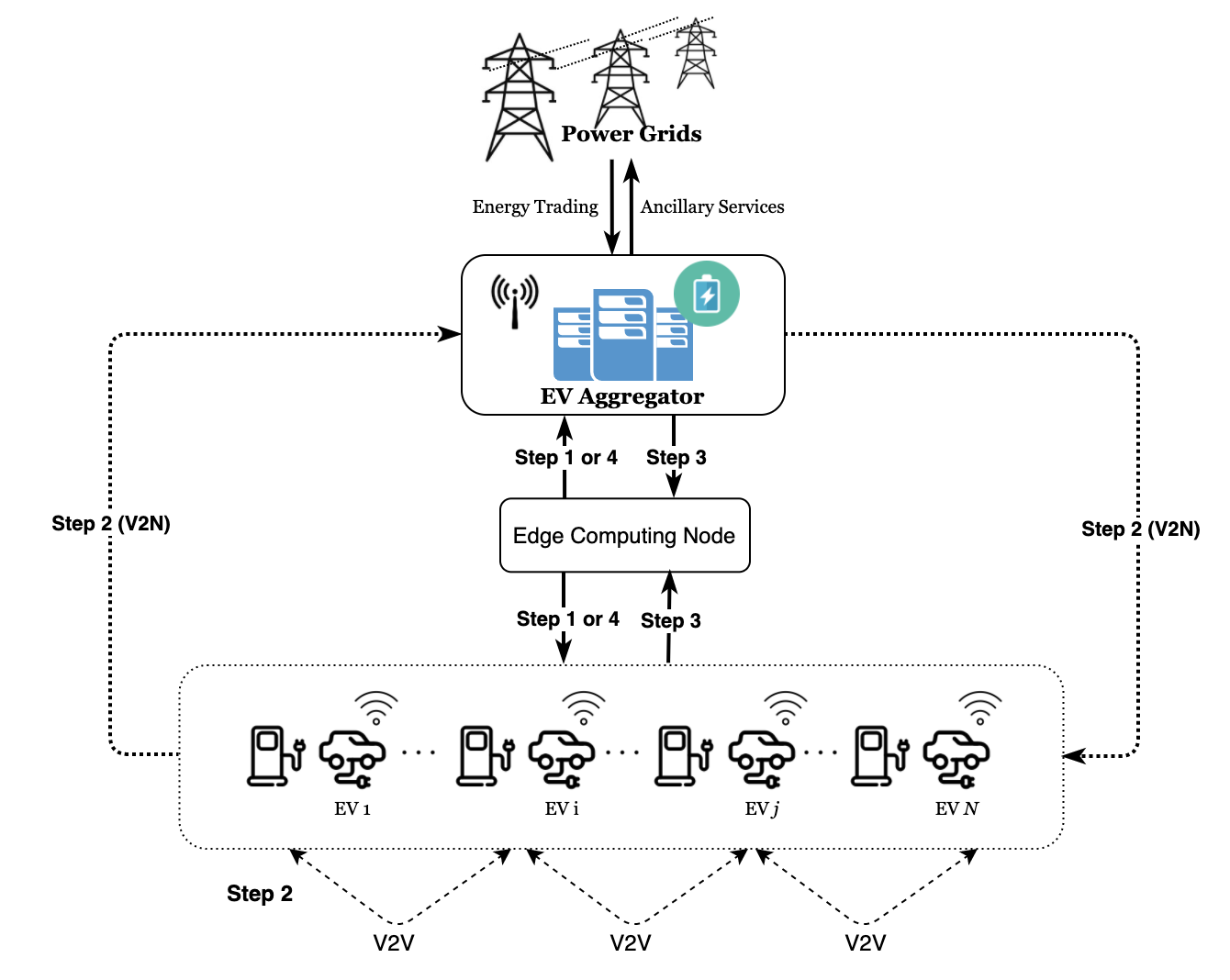}}
		\caption{A schematic diagram of the V2G scenario using the DWOA. Note that V2V indicates the communication between EVs, and V2N indicates the communication between vehicles and the EV aggregator (Network). Both communications are required in Step 2. }
		\label{schematic}
	\end{center}
\end{figure*}

In Section \ref{WOA}, we have seen that there are two challenges for using CWOA in our system design, namely 1. dealing with the hard constraint due to the consensus requirement; and 2.  the privacy-preserving mechanism for both EVs and the EV aggregator. In this section, we borrow the fundamental ideas from CWOA and devise the Decentralized WOA (DWOA) to better tackle with the two challenges. To implement DWOA, our proposed system architecture is illustrated in Fig. \ref{schematic}, which includes three main components, namely an EV aggregator, an edge computing node and a group of EVs using V2G. Based on this architecture, our system operates in four steps. 

\subsection*{Step 1: Initialization}

In this step, an edge computing node (ECN) initializes $M$ whales in a sequence. Each whale essentially represents a potential optimal discharge rate to be evaluated by each EV at the later phase. When an initial (random) sequence of numbers has been generated, the ECN broadcast this sequence to both EVs and the EV aggregator through proper communication channels, e.g. WiFi or power-line communication. 

\subsection*{Step 2: Local Evaluation, Data Shuffling and Aggregation}

For each EV $i$, let $c_i^h(k)$ be the value of the $h$'th element at $k$'th iteration in the received sequence of length $M$. Similarly, we use the same notation for the same copy of data sequence received by the EV aggregator.  Given the unit price $p$, each EV evaluates $f_i(c_i^h(k);p), \forall i, h$ and the EV aggregator evaluates $Agg(\sum_i c_i^h(k)), \forall h$. This finishes the local evaluation part in Step 2. However, these locally evaluated data values cannot be sent directly to ECN for further processing as it violates the privacy-preserving requirement in our system design. To deal with this concern, we propose a data shuffling and aggregation procedure to allow privacy-preserving computing for these values at the ECN. For this purpose, let us denote $\underline{\textrm{N}}_i(k)$ the set of neighbours of the $i$'th EV at $k$'th iteration. In this regard, please note that the EV aggregator can always be seen as one of the neighbours in $\underline{\textrm{N}}_i(k)$ for the $i$'th EV. Similarly, the available EVs are also seen as neighbours of the EV aggregator which can be described using the set $\underline{\textrm{N}}_a(k)$. It is worth mentioning that the neighbouring set defines those agents which can be connected by the agent of interest, i.e. it is the outdegree rather than the indegree of a node (vertex) in a corresponding directed graph for communication links. With this in mind, the algorithm to implement this proposed procedure is demonstrated in Algorithm \ref{alg:step2}. 

 \begin{algorithm}[htbp]
 	\caption{Data Shuffling and Local Aggregation}
 	\begin{algorithmic}[1]
 		\For{each EV $i \in \underbar{N}(t)$ }
 		\State Splits each mapping $(c_i^h(k), f_i(c_i^h(k);p)), \forall h$, into two copies.
 		\State Sends one copy to anyone of its neighbours in $\underline{\textrm{N}}_i(k)$.
 		\State Reserves one copy locally. 
 		\State Aggregates received copies from other neighbours to construct a new mapping  $(c_i^h(k), f_i^{*}(c_i^h(k);p))$
 		\State Sends the new mapping $(c_i^h(k), f_i^{*}(c_i^h(k);p))$ to ECN for further processing. 
 		\EndFor \\
 		
 		\For{the EV aggregator}
 		\State Splits each mapping $(c_i^h(k), Agg(\sum_i  c_i^h(k))), \forall h$, into two copies.
 		\State Sends one copy to anyone of its neighbours in $\underline{\textrm{N}}_a(k)$.
 		\State Reserves one copy locally. 
 		\State Aggregates received copies from other neighbours to construct a new mapping  $(c_i^h(k), Agg^{*}(\sum_i  c_i^h(k)))$. 
 		\State Sends the new mapping $(c_i^h(k), Agg^{*}(\sum_i  c_i^h(k)))$ to ECN for further processing. 
 		\EndFor
 	\end{algorithmic}
 	\label{alg:step2}
 \end{algorithm}

For EVs, the splitting operation essentially divides the value of $f_i(c_i^h(k);p), \forall i, h$ into two parts in an arbitrary manner. This splitting method is valid only if the two divided values add up the same as the original one. Then, one copy will be sent to a neighbouring EV/aggregator which ensures that the local aggregated value for the mapping will be different, i.e. $f_i^{*}(c_i^h(k);p) \neq f_i(c_i^h(k);p)$ as the $i$'th EV also receives different copies from other neighbours which will be used for its local aggregation for $f_i^{*}(c_i^h(k);p)$. Similar idea also applies for the EV aggregator. It is this procedure which makes the data flow privacy-aware for different parties involved in the computation. In particular, the ECN is not able to identify $f_i$ or $Agg$ from the received data flow. However, as we shall see later, ECN will use this information to steer the search agent to a better solution by leveraging a WOA-based approach. To better illustrate this idea, we now give a numerical example. \\

\textit{Example:} We assume that for the EV $i$, there are two mapping pairs that have been evaluated, namely (1, 5) and (2, 10). Also, we assume that for the EV $j$, there are also two mapping pairs that have been evaluated, i.e. (1, 7), (2, 20). The second value in each mapping pair is different as it depends on the EV's own cost function in our case. We assume that EV $i$ is the neighbour of the EV $j$, and vice versa. We assume that the values will be split in the following manner:  

\begin{itemize}
	\item (1, 5) $\rightarrow$ (1, 2) and (1, 3); (2, 10)  $\rightarrow$ (2, 3) and (2, 7)
	\item (1, 7) $\rightarrow$ (1, 5) and (1, 2); (2, 20) $\rightarrow$ (2, 5) and (2, 15).
\end{itemize}

Next, we assume that EV $i$ will send one arbitrary copy of each mapping to EV $j$, and so does EV $j$. After this swapping, the new local copies from both EVs are reported as follows:

\begin{itemize}
	\item (1, 5) $\rightarrow$ (1, 2) and (1, 5); (2, 10)  $\rightarrow$ (2, 15) and (2, 7)
	\item (1, 7) $\rightarrow$ (1, 3) and (1, 2); (2, 20) $\rightarrow$ (2, 5) and (2, 3). 
\end{itemize}

After local aggregation, we can obtain the new mappings for both EVs as follows:

\begin{itemize}
	\item EV $i$: (1, 5) $\mapsto$ (1, 7);  (2, 10) $\mapsto$ (2, 22)
	\item EV $j$: (1, 7) $\mapsto$ (1, 5);  (2, 20) $\mapsto$ (2, 8)
\end{itemize}
 
Finally, we note that although the splitting and sending procedures are fully arbitrary, the total sum for all mappings related to a specific key value is equivalent. For example, before Step 2, the total sum for the discharge rate 1 is 12, and this result is consistent after Step 2.  

\subsection*{Step 3: ECN Aggregation, DWOA and Decision Making}

After finishing Step 2, ECN has received all local aggregated copies from EVs and the EV aggregator. Next, the ECN is required to determine which discharge rate is more preferable for the optimization problem \eqref{eq:opt} given all mappings received. Considering the objective function defined in \eqref{eq:opt}, ECN carries out another local aggregation to find out which discharge rate results in the minimal total mapped value. Taking the example above, ECN will select discharge rate 1 instead of 2 to proceed simply because discharge rate 1 has a total mapped value 12 but discharge rate 2 has a total mapped value 30. 

After the aggregation at ECN, the most preferable discharge rate from current configurations has been selected. However, there is still no guarantee that the current best solution is the best solution among all other discharge rates. To address this concern, we now present the DWOA in Algorithm \ref{woa_new}, where $\alpha$ is a scalar linearly decreased from 2 to 0, and $r$ is a random number between $[0, 1]$.

\begin{algorithm}[htbp]
	\caption{Decentralized Whale Optimization Algorithm}
	\begin{algorithmic}[1]
		\If{$k < k_{\textrm{max}}$}	
		\For{h = 1, 2, \dots, M}
		\State Update $A = 2\alpha r - \alpha$.
		\State Update $C = 2 r$.
		\State Update random numbers $l \in [-1, 1]$ and $p \in [0,1]$.
		\If{$p < 0.5$}
		\If{$|A| < 1$}
		\State $c_i^{h}(k+1) = c_i^{h^{*}}(k) - A ~| C \cdot c_i^{h^{*}}(k) -  c_i^{h}(k) | $
		\ElsIf{$|A| \geq 1$} 
		\State  Select a random $c_i^{\tilde{h}}(k)$.
		\State  $c_i^{h}(k+1) = c_i^{\tilde{h}}(k) - A ~| C \cdot c_i^{\tilde{h}}(k) -  c_i^{h}(k) | $
		\EndIf           
		\ElsIf{$p \geq 0.5$}
		\State Calculate $d_i =  | c_i^{h^{*}}(k)  -  c_i^{h}(k)| $
		\State $c_i^{h}(k+1) =   d_i \cdot e^{l} \cdot cos(2\pi l) + c_i^{h^{*}}(k)$ 
		\EndIf
		\EndFor
		\State  Check if $c_i^{h}(k+1), \forall h$ goes beyond the search space and amend it.
		\State  Return $c_i^{h}(k+1), \forall i \in \underline{\textrm{N}}(t)$. 
		\State  $k = k + 1$
		\State  Proceed to Step 4. 
		\Else  
		\State  Return $c_{i}^{h^{*}}(k_{\textrm{max}}), \forall i \in \underline{\textrm{N}}(t)$, and stop the calculation.
		\EndIf
	\end{algorithmic}
	\label{woa_new}
\end{algorithm}

\subsection*{Step 4: Updating Parameters for Next Iteration}

In Step 3, the ECN has obtained the best current solution from local aggregated results, and the DWOA is used to improve the current solution and outputs the adjusted solution for a potentially better solution. In the final step, i.e. Step 4, the ECN will broadcast this updated solution to all EVs and the EV aggregator for the next iteration. After this broadcast, all EVs and the EV aggregator will evaluate their new mappings and start the loop from the Step 2 again until the simulation terminates (i.e. maximum number of iterations has been reached). \\

\noindent \textbf{Remark:} The key idea of the proposed system implementation is to set up a mechanism so that different parties, i.e. EVs and the aggregator, can be involved in such a joint calculation without violating any privacy constraint. Roughly speaking, the aggregation step at ECN can be seen as an evaluation of the fitness function in the CWOA to determine which solution is the ``lead whale''. ECN implements the proposed DWOA to steer other solutions (``non-lead whales'') towards better solutions (``hunting positions'') given the current best solution (``the lead whale position'').  Finally, the broadcast mechanism at the ECN preserves the consensus for the optimal solution. We shall expect that after maximum number of iterations, $k_{\textrm{max}}$, the final value $c_{i}^{h^{*}}(k_{\textrm{max}})$ can be used as the optimal solution for \eqref{eq:opt} at a given slot $t$.

\subsection{System Interfaces}

We note that it is not the main focus of this paper to devise various practical interfaces for the real-world implementation of the proposed system, however we find it would be useful to provide some general observations to facilitate the operation and deployment of such a system in reality.

\begin{itemize}
	\item EV users: when an EV arrives at the parking area, the user needs to specify a minimal acceptable state-of-charge value $SOC_\textrm{min}^i$ to be used for the V2G programme. 
	\item EVs: each EV needs to have an onboard software tool which is able to evaluate/retrieve its cost value, splitting cost mappings and conduct local aggregation procedures required in Step 2. Each EV is required to have a communication capability with its connected V2G discharging point through mutually acceptable V2G communication standards/protocols. 
	\item V2G discharging points: the communication/computing unit of each discharging point needs to exchange messages with other discharging points which are managed by the EV aggregator. The unit is also able to exchange messages with the ECN and the central server at the EV aggregator. The unit can communicate the real-time pricing signal and the current discharge rate information to the connected EV. The unit should also know the energy conversion efficiency for the connected EV, i.e. $\eta_i$, and this can be easily estimated from the DC/AC power.
	\item ECN: the unit can be a smart gateway or a server located closer to the discharging points in a real-world scenario. This unit is required to conduct all the computing and communication tasks in Steps 1, 3 and 4. 
	\item EV aggregator: similar to what is required for EVs in Step 2, the central server at the EV aggregator is required to evaluate/retrieve its cost value, split cost mappings and conduct local aggregation procedures given the discharge rates received from the ECN. The server also communicates to EVs through the discharging points.
\end{itemize}

\section{Simulations} \label{Simulation}

In this section we introduce our simulation set-up in Matlab and then we evaluate the performance of our proposed system implementations.

\subsection{Simulation Setup}\label{setup}

\begin{figure}[htbp]
	\begin{center}
		{\includegraphics[scale=0.4]{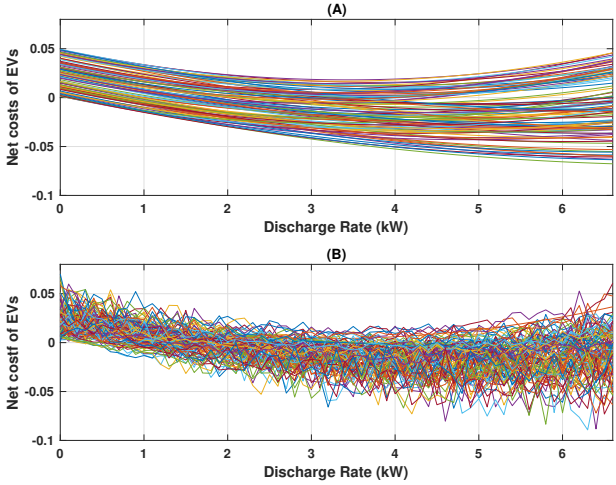}}
		\caption{Cost functions of 100 EVs used in the simulation before Step 2 (A) and after Step 2 (B). }
		\label{cost_function_evs}
	\end{center}
\end{figure}

\begin{figure}[htbp]
	\begin{center}
		{\includegraphics[scale=0.4]{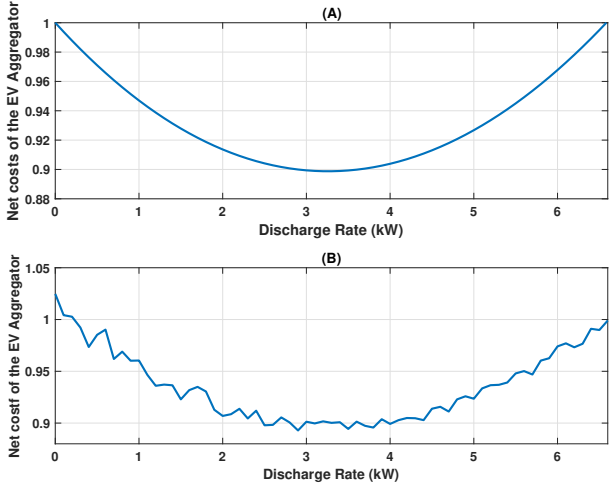}}
		\caption{The cost function of the EV aggregator used in the simulation before Step 2 (A) and after Step 2 (B). Note that we only evaluate the case of our interest in which the discharge rates of all EVs are consistent.}
		\label{cost_function_ev_agg}
	\end{center}
\end{figure}

We consider a simple scenario with 100 EVs connected to an EV aggregator for energy trading using V2G. We assume that the initial state-of-charge of a group of EVs is uniformly distributed between 80\% and 90\% upon arrival, with a minimal state-of-charge uniformly distributed between 10\% and 20\%. For a given EV, we assume that its battery size is uniformly distributed between 15 to 30 kWh, which is consistent with most commercial EVs in the market. Moreover, we specify the minimal discharge rate $c_\textrm{min}^i = 0$kW, and the maximum discharge rate $c_\textrm{max}^i = 6.6$kW which is consistent with the AC Level 2 standard connector in \cite{botsford2009fast}. For a given price signal $p=0.02$, we set the parameters for both EVs and the EV aggregator to obtain the cost functions for our simulations. For simplicity of our system model, we assume that each EV has only one neighbour, either an EV or the EV aggregator. In Fig. \ref{cost_function_evs}, we show the plot (A). original cost functions for EVs; and (B). the altered (locally aggregated) cost functions for EVs after one iteration of the Step 2.  Likewise, in Fig. \ref{cost_function_ev_agg} we show the original cost function for the EV aggregator in the subplot (A) and the altered cost function after the Step 2 in (B). It is clear that all functions have been altered for the privacy-preserving computing. 

\subsection{Simulation Results}

\begin{figure}[htbp]
	\begin{center}
		{\includegraphics[scale=0.36]{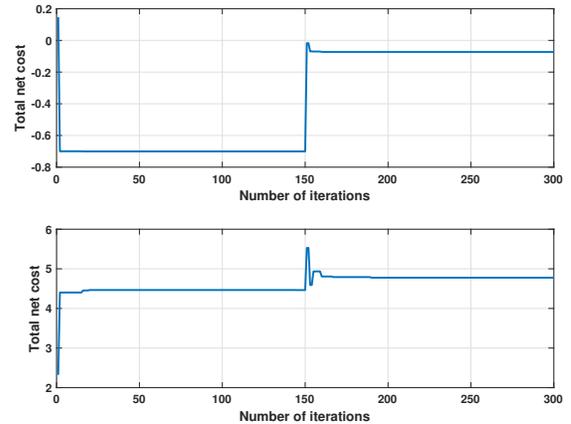}}
		\caption{Converged results for the DWOA solution for 100 EVs (within 0-150 iterations) and 50 EVs (after 150 iterations).}
		\label{agg_converge}
	\end{center}
\end{figure}

To begin with, we shall evaluate the dynamic performance of the proposed system. For this purpose, we first assume that 100 EVs are connected to the V2G discharging points at the start of the simulation, and the DWOA will be implemented to calculate the optimal consensus discharge rate for the 100 EVs within 150 algorithm iterations. After that, let 50 EVs still be connected to the discharge points, so that the system will be adapted to recalculate the new optimal solution for the EVs. By default, we assume that there is only one whale involved in the calculation in this case, i.e. $M=1$. The simulation result is shown in Fig. \ref{agg_converge}. This result demonstrates that the DWOA can very efficiently capture the rapid change of EVs connected to the V2G points. Specifically, the first optimal solution has been found in less than 25 iterations to 4.46 kW, and the new optimal solution for 50 EVs has been converged to 4.78 kW within just 30 iterations of the algorithm. 

\begin{figure}[htbp]
	\begin{center}
		{\includegraphics[scale=0.36]{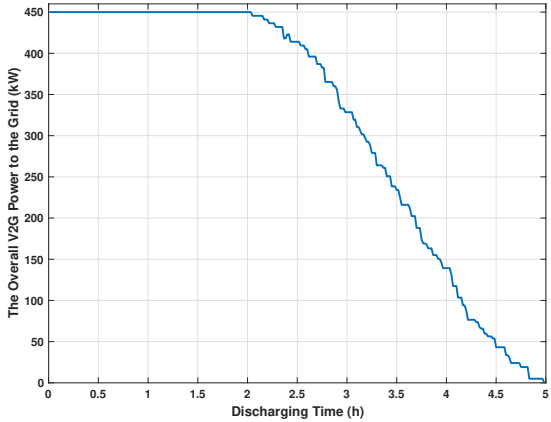}}
		\caption{V2G Power delivered to the grid while EVs are available.}
		\label{grid_result1}
	\end{center}
\end{figure}

\begin{figure}[htbp]
	\begin{center}
		{\includegraphics[scale=0.36]{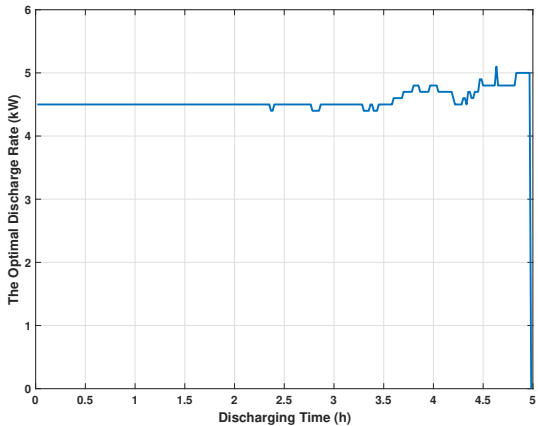}}
		\caption{Evolution of the optimal discharge rate for EVs during V2G.}
		\label{grid_result2}
	\end{center}
\end{figure}

\begin{figure}[htbp]
	\begin{center}
		{\includegraphics[scale=0.36]{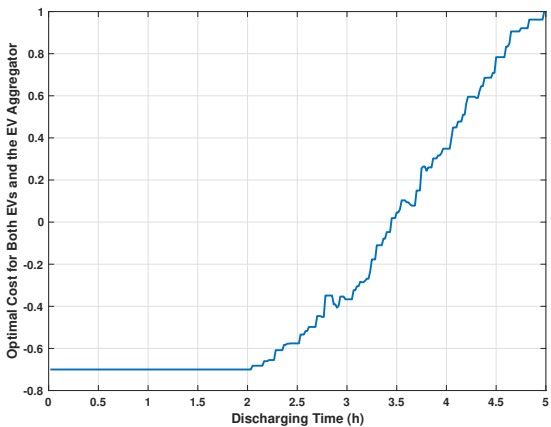}}
		\caption{Evolution of the optimal cost when EVs set to the optimal discharge rate during V2G. }
		\label{grid_result3}
	\end{center}
\end{figure}

\begin{figure}[htbp]
	\begin{center}
		{\includegraphics[scale=0.36]{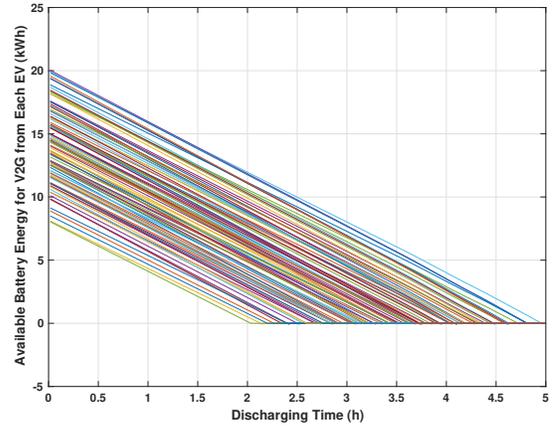}}
		\caption{Evolution of the available energy for EVs during V2G.}
		\label{grid_result4}
	\end{center}
\end{figure}

In Figs. \ref{grid_result1} - \ref{grid_result4}, we illustrate that the efficacy of the proposed system from the perspective of energy management. In Fig. \ref{grid_result1}, it shows the overall V2G power that can be provided to the grid during the time when EVs are available, i.e. $SOC_i(t) > SOC_{\textrm{min}}^i$. Due to the unavailability of some EVs over time, the total available power from EVs to the grid decreases as expected. It is worth noting that with 100 EVs plugged in for V2G, the EV aggregator can supply a peak power of 450 kW for the grid demand. In Fig. \ref{grid_result2}, it shows how the optimal discharge rate will be evolved during this process by using DWOA. We have found that when EVs gradually ``leave out'' the discharge points, the optimal discharge rates will be changed accordingly to capture the benefits for the remaining EVs which still use V2G. Here, we also note that the obtained optimal discharge rate is significantly different compared to optimizing the benefits for one party only, e.g. the optimal discharge rate for only the EV aggregator is around 3.2 kW as shown in Fig. \ref{cost_function_ev_agg}.

\begin{figure}[htbp]
	\begin{center}
		{\includegraphics[scale=0.36]{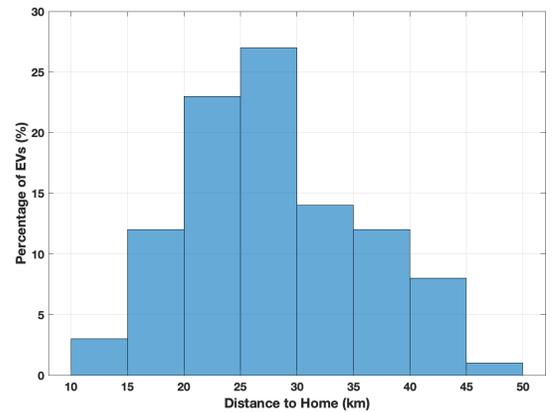}}
		\caption{Estimated distance to home according to the initial SOC specified by the EV users.}
		\label{home_distance}
	\end{center}
\end{figure}

Further, Fig. \ref{grid_result3} shows how the optimal cost can be changed accordingly when the optimal discharge rate has been applied in Fig. \ref{grid_result2}. As expected again, the overall optimal cost increases when EVs gradually  become unavailable, i.e. the EV aggregator now acts as the dominant factor, which has an optimal cost around 0.9 reflected in Fig. \ref{cost_function_ev_agg}. In Fig. \ref{grid_result4} it shows how the available energy evolves for each EV in V2G. Clearly, it shows that the discharge rate is indeed the same for a group of EVs which are currently available at the discharge points. It also shows that EVs will immediately become unavailable when its current state of charge is less than what it initially specified. Finally, to further illustrate how the initial $SOC$ can affect EV users' travelling distance back home, we plot our simulation result in Fig. \ref{home_distance}. We assume that the distance per kWh is 8.26 km according to \cite{mruzek2016analysis} and the resulting histogram shows that with current configurations for $SOC_{\textrm{min}}^i$, most EV users (45\%) can have a home distance between 20 - 30km without any concern on the energy depletion back home. This will largely motivate local EV users to use the proposed V2G programme.

\begin{figure}[htbp]
	\begin{center}
		{\includegraphics[scale=0.36]{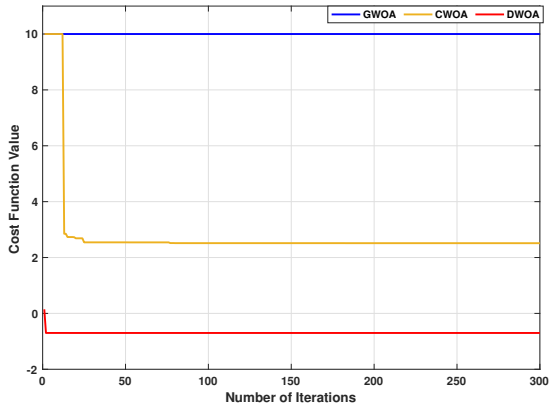}}
		\caption{Comparison of simulation results for three optimization algorithms.}
		\label{gwo_compare}
	\end{center}
\end{figure}

To further illustrate the efficacy of the DWOA, we have also compared the dynamic performance of the DWOA with the CWOA and the Grey Wolf Optimization Algorithm (GWOA) \cite{mirjalili2014grey} as our baseline algorithms. We note that in order to implement both CWOA and GWOA, a penalty item has to be introduced for the evaluation of the fitness function. In this case, we set the maximum penalty for violating the consensus constraint being 10. The simulation result is shown in Fig. \ref{gwo_compare}. Clearly, the proposed DWOA can effectively converge to optimality with a very limited number of iterations. In contrast, the other two baseline algorithms cannot converge to the optimal solution within 300 iterations due to the hard consensus constraint. However, we do observe that CWOA achieves a better result compared to GWOA as in the latter case the algorithm has been stuck in dealing with the constraint and can hardly evolve. 

Since  the DWOA is also a heuristic algorithm, it is useful to discuss its statistical performance as well. Specifically, we shall investigate how the algorithm can perform with changes in different hyper-parameters, such as the number of whales parameter $M$ and maximum number of iterations $k_{\textrm{max}}$. For this purpose, we applied the default setup of cost functions as aforementioned, and then conducted 10000 independent algorithm experiments for each scenario of our interest. \footnote{We used Matlab R2019b on a Macbook Pro (macOS Version 10.14.6) with 2.3GHz Intel Core i9 and 16GB 2400 MHz DDR4 Memory.} First, we progressively increased $k_{\textrm{max}}$ from 50 to 400, whilst fixing the number of initialized whales $M$ of each EV to 1. The corresponding simulation results are included in Table \ref{sim_result_t1}.  The results show that with increasing value of $k_\textrm{max}$ the mean value of the converged discharge rate remains almost identical for all cases, and with nearly negligible value of the standard derivation shown in the third column. The last column in the table lists the actual running time of the system of all independent experiments in our simulation environment. This can be used as an estimator for further evaluation of system time complexity. Clearly, the system running time increases with increasing value of $k_\textrm{max}$. However, the parameter $k_\textrm{max}$ shows very little sensitivity in terms of accuracy, i.e. the mean value of the final converged discharge rates in all runs (Mean DR) and its standard derivation (Std DR).

\begin{table}[htbp]
	\caption{Simulation Results by Changing $k_{\textrm{max}}$ while fixing $M=1$.} 
	\begin{center}
		\begin{tabular}{|l|c|c|c|}
			\hline
			Max Iteration & Mean DR (kW)& Std DR (kW)& Time (s) \\
			\hline
		    50  & 4.4627 &  5.2e-4 &  6.72 \\
			100 &  4.4628 & 1.27e-5  & 12.67 \\
			150 & 4.4628 & 1.60e-5  & 18.83  \\
			200 & 4.4628 & 3.19e-5 & 24.87 \\
			250 & 4.4628 & 1.79e-6 & 30.13 \\
			300 & 4.4628 & 4.13e-6 & 35.87 \\
			350 & 4.4628 & 3.47e-6 & 42.01 \\
			400 & 4.4628 & 7.05e-7 & 47.98 \\
			\hline
		\end{tabular}
	\end{center}
	\label{sim_result_t1}
\end{table} 

Finally, we investigate a scenario by varying the number of whales $M$ while fixing $k_\textrm{max} $ to 300. These results are presented in Table \ref{sim_result_t2}. It is clear that both Mean DR and the Std DR have been almost consistent in all cases with negligible value of Std DR. However, the overall computation time has been significantly increased compared to the results in Table \ref{sim_result_t1} due to the fact that more whales need to be updated in each algorithm iteration. \\

\begin{table}[htbp]
	\caption{Simulation Results by Changing $M$ while fixing $k_{\textrm{max}}=300$. } 
	\begin{center}
		\begin{tabular}{|l|c|c|c| }
			\hline
			Number of Whales & Mean DR (kW) & Std DR (kW)& Time (s)\\
			\hline
			1  & 4.4628 &  4.13e-6  & 35.87 \\
			10 & 4.4628& 1.02e-7 & 270.53 \\
			20 & 4.4628 & 3.43e-9  & 600.22\\
			30 & 4.4628 & 2.18e-8  & 816.58\\
			40 & 4.4628 & 5.49e-9  & 1038.27\\
			\hline
		\end{tabular}
	\end{center}
	\label{sim_result_t2}
\end{table}

\section{Conclusion}\label{conclusion}

In this paper, our overachieving goal is to minimize the overall net costs for both EVs and the EV aggregator in a privacy-preserving manner. To do this, we modified the WOA to its decentralised form, DWOA, to solve an optimal consensus problem while leveraging simple ideas of data shuffling and aggregation to address the privacy concerns when multi-users need to be involved in the joint computing process. Many simulation results have been included to demonstrate the efficacy of the proposed system implementation. Specifically, we have evaluated the dynamical performance of DWOA and how it performs on the grid side. We have also presented results on how the proposed algorithm compares with other two baseline algorithms, and our results have shown that the DWOA can converge to optimality very efficiently compared to others. Finally, we have shown some statistical results to further illustrate the robustness of our proposed system in solving the challenging V2G problem. In particular, we have shown that the DWOA can converge to optimality within 50 algorithm iterations, which can almost be done instantly in a real-world V2G scenario referring to the little computation time revealed in our simulation results.

\section{Limitations and Future Extensions}\label{future}

The work presented in the paper is an important step towards ``a privacy-preserving system design of a fair V2G programme involved both EVs and the EV aggregator''. This is only a first step of work and it neglects some aspects of a complete solution, which will be the key topics in our future work. For instance, the proposed system does not model the complex time-varying evolution of communication links in the V2G network, as well as how it affects the optimal solution and the system implementation. Finally, we assumed that the EV users in the V2G programme are mostly interested in selling their battery power to the grid, and we did not consider the charging aspect which a user may also need in the V2G programme, which may be investigated in future work as well.


\bibliographystyle{IEEEtran}
\bibliography{References}

\begin{IEEEbiography}[{\includegraphics[width=1in,height=1.25in,clip,keepaspectratio]{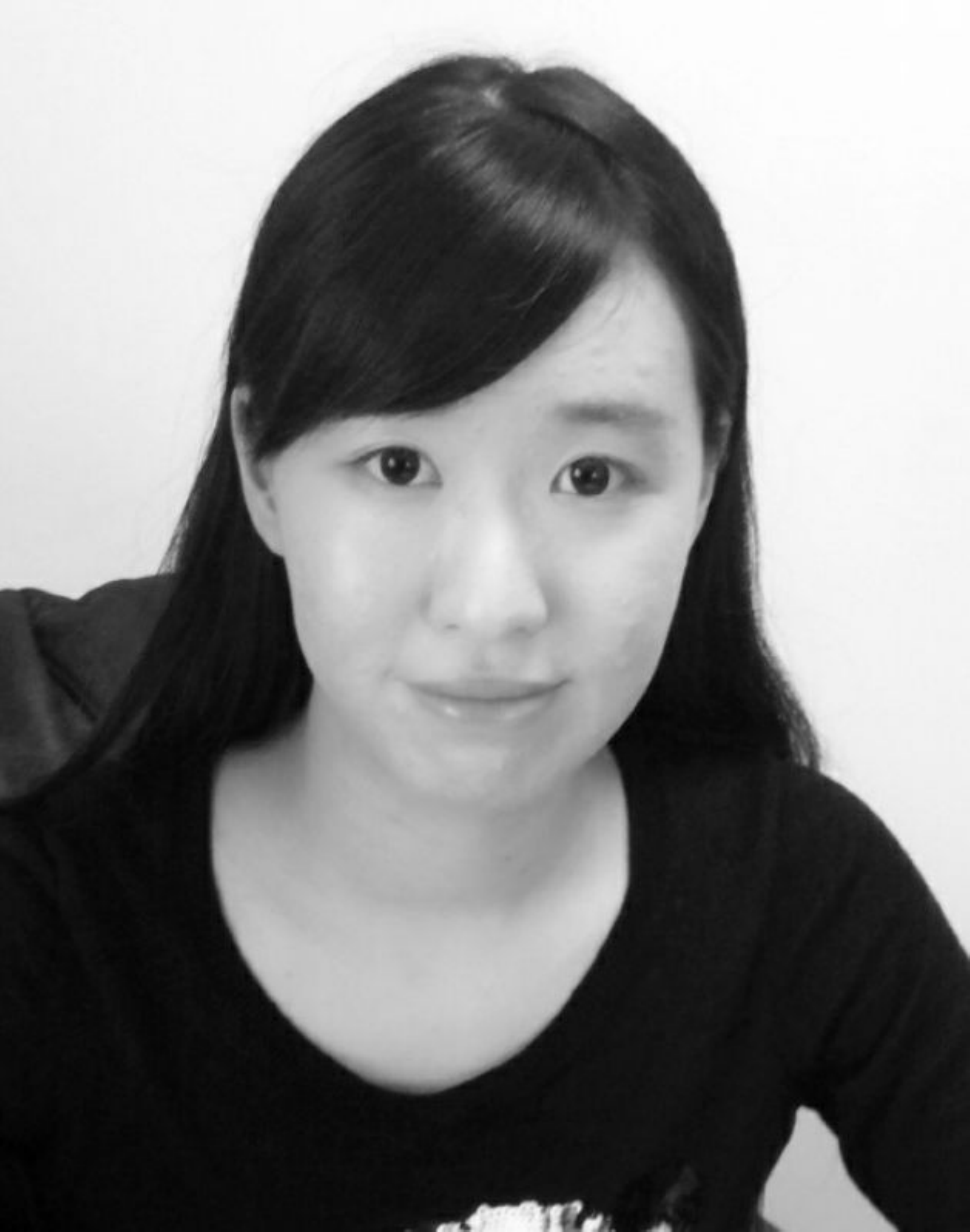}}]{Yingqi Gu} is currently a data scientist at Alkermes plc. She was a postdoc research fellow with the Insight Centre for Data Analytics at Dublin City University and the adjunct lecturer at Maynooth University. She received her Master of Science degree from School of Engineering, the University of Edinburgh and the Ph.D. degree in Control Engineering and Decision Science from School of Electrical and Electronic Engineering, University College Dublin. Her current research interests include AI \& machine learning, control and optimisation theories with applications to EVs, intelligent transportation systems, and digital health.
\end{IEEEbiography}

\begin{IEEEbiography}[{\includegraphics[width=1in,height=1.25in,clip,keepaspectratio]{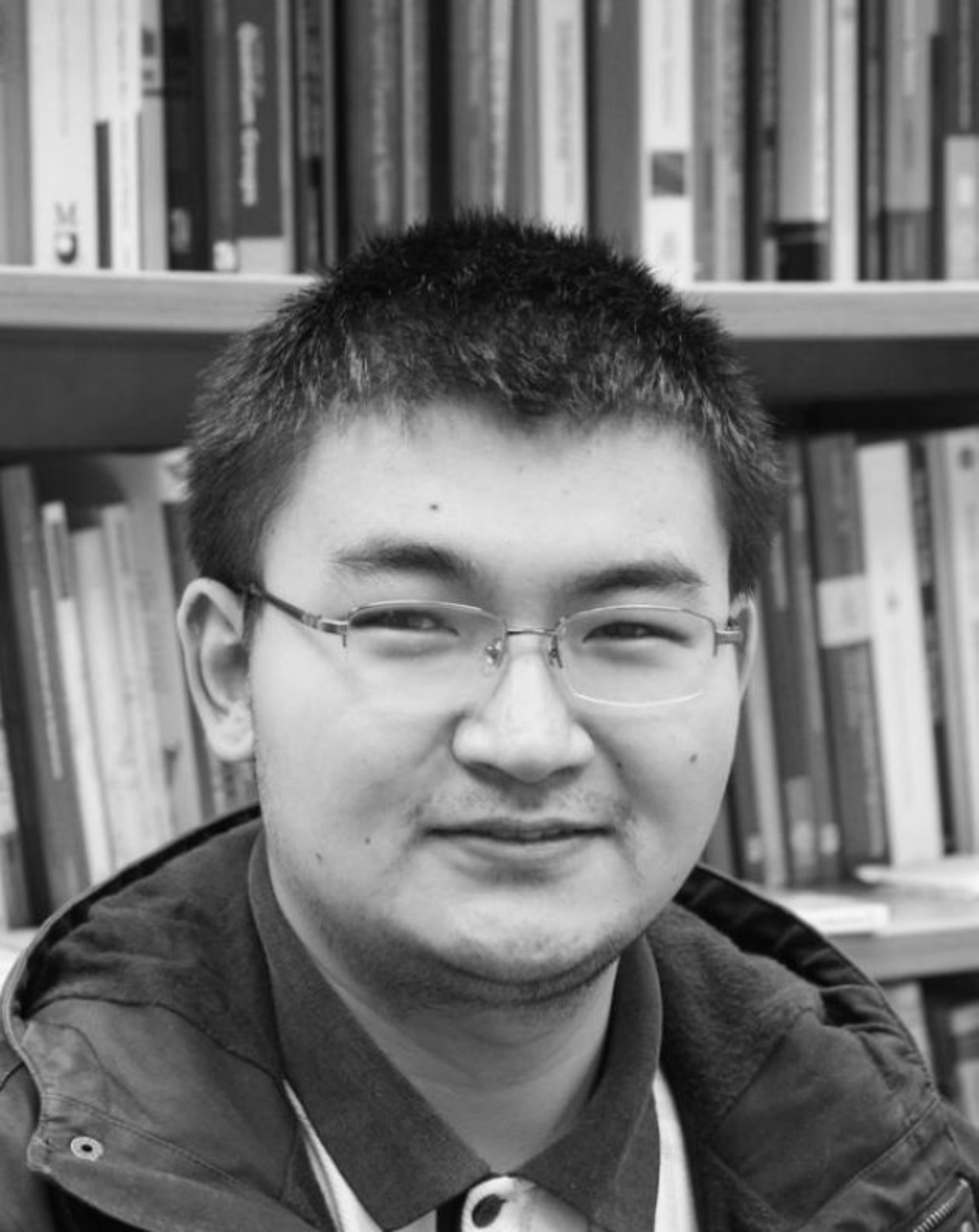}}]{Mingming~Liu}
received his B.E. degree in Electronic Engineering with first class honours from National University of Ireland Maynooth in 2011, and his Ph.D. degree from the Hamilton Institute, Maynooth University in 2015. After that, he worked at University College Dublin and IBM Ireland Lab as an applied researcher, data scientist and EU H2020 project lead. Since 2019, he has been working as an Assistant Professor in the School of Electronic Engineering at Dublin City University. His research interests include control and optimization theories with applications to 5G \& IoT, electric vehicles, smart grids, intelligent transportation systems and smart cities. 
\end{IEEEbiography}

\end{document}